\documentclass[sigconf]{acmart}

\copyrightyear{2025}
\acmYear{2025}
\setcopyright{cc}
\setcctype{by-nc-sa}
\acmConference[DIS '25 Companion]{Designing Interactive Systems Conference}{July 5--9, 2025}{Funchal, Portugal}
\acmBooktitle{Designing Interactive Systems Conference (DIS '25 Companion), July 5--9, 2025, Funchal, Portugal}
\acmDOI{10.1145/3715668.3736363}
\acmISBN{979-8-4007-1486-3/2025/07}

\usepackage{color}
\usepackage{soul}    % highlighting
\usepackage{float}

\usepackage{verbatim}
\usepackage{listings} % in the preamble

\usepackage{setspace}

\graphicspath{ {figures/} }
 
\usepackage{gensymb} % degree symbol

\usepackage{enumitem} % remove indent of list items

\usepackage{xspace} % for putting space when necessary

\usepackage{graphicx}

\newcommand{\systemName}{\textsc{ImprovMate}\xspace}

\begin{document}

\title{\systemName: Multimodal AI Assistant for Improv Actor Training}

\author{Riccardo Drago}
\orcid{0009-0009-3628-044X}
\affiliation{%
  \institution{Politecnico di Torino}
  \city{Torino}
  \country{Italy}
}
\email{riccardo.drago@studenti.polito.it}

\author{Yotam Sechayk}
\orcid{0009-0002-5286-0080}
\affiliation{%
  \institution{The University of Tokyo}
  \city{Tokyo}
  \country{Japan}
}
\email{ysechayk@acm.org}

\author{Mustafa Doga Dogan}
\orcid{0000-0003-3983-1955}
\affiliation{%
  \institution{Adobe Research}
  \city{Basel}
  \country{Switzerland}
}
\email{doga@adobe.com}

\author{Andrea Sanna}
\orcid{0000-0001-7916-1699}
\affiliation{%
  \institution{Politecnico di Torino}
  \city{Torino}
  \country{Italy}
}
\email{andrea.sanna@polito.it}

\author{Takeo Igarashi}
\orcid{0000-0002-5495-6441}
\affiliation{%
  \institution{The University of Tokyo}
  \city{Tokyo}
  \country{Japan}
}
\email{takeo@acm.org}

%%
%% The "author" command and its associated commands are used to define
%% the authors and their affiliations.
%% Of note is the shared affiliation of the first two authors, and the
%% "authornote" and "authornotemark" commands
%% used to denote shared contribution to the research.

%%
%% By default, the full list of authors will be used in the page
%% headers. Often, this list is too long, and will overlap
%% other information printed in the page headers. This command allows
%% the author to define a more concise list
%% of authors' names for this purpose.

\renewcommand{\shortauthors}{Drago, et al.}

\begin{abstract}

Improvisation training for actors presents unique challenges, particularly in maintaining narrative coherence and managing cognitive load during performances. Previous research on AI in improvisation performance often predates advances in large language models (LLMs) and relies on human intervention. We introduce \systemName, which leverages LLMs as GPTs to automate the generation of narrative stimuli and cues, allowing actors to focus on creativity without keeping track of plot or character continuity. Based on insights from professional improvisers, \systemName incorporates exercises that mimic live training, such as abrupt story resolution and reactive thinking exercises, while maintaining coherence via reference tables. By balancing randomness and structured guidance, \systemName provides a groundbreaking tool for improv training.
Our pilot study revealed that actors might embrace AI techniques if the latter mirrors traditional practices, and appreciate the fresh twist introduced by our approach with the AI-generated cues.

\end{abstract}

%%
%% The code below is generated by the tool at http://dl.acm.org/ccs.cfm.
%% Please copy and paste the code instead of the example below.
%%


%%
%% Keywords. The author(s) should pick words that accurately describe
%% the work being presented. Separate the keywords with commas.

%% A "teaser" image appears between the author and affiliation
%% information and the body of the document, and typically spans the
%% page.
\begin{teaserfigure}
  \centering
  \includegraphics[width=1\textwidth]{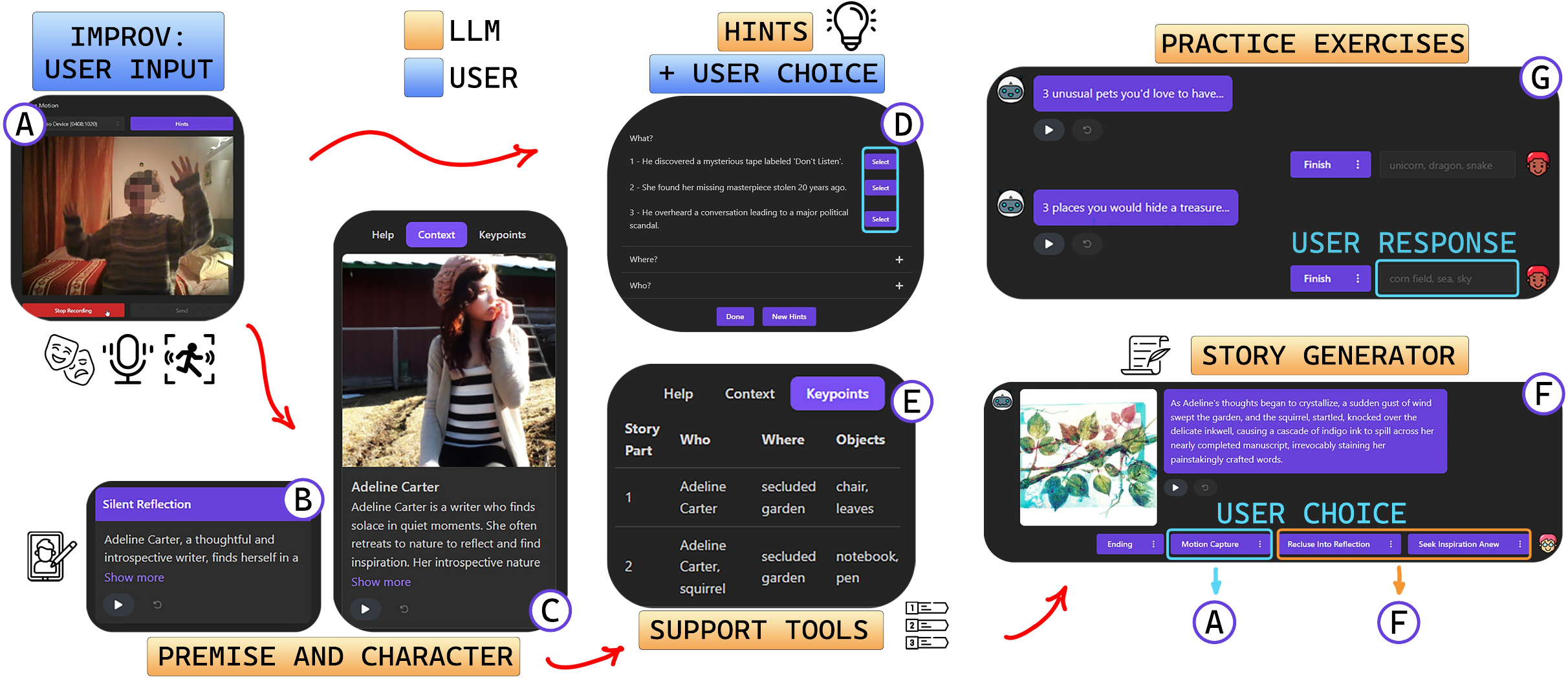}
  \caption{The core interaction loop within \systemName. An actor uses full-body movements and dialogue to stage an improvisation: (A) motion input, (B) the premise for the AI-generated story, (C) AI-generated character based
  on the context given by the user (\textit{image AI-generated}),  (D) AI-generated suggestions to help improvise, (E) the tool to track key points in the story, (F) AI-generated story consistent with the performance, advancing the narrative, (G) an exercise to practice and improve the actor's responsiveness. %Key components of the motion capture and AI feedback mechanisms are highlighted to depict the integration of user inputs into the improvisation experience.
  }
  \Description{ImprovMate interaction loop schematic. A left-to-right diagram of seven labeled components showing how an improvising actor's motions and speech drive AI support: (A) user input: full-body motion capture, facial tracking and microphone icons beneath a performer image; (B) premise: an AI-generated story premise, with play and pause buttons alongside a ``Show more'' link; (C) character: a profile card for the character with an AI-generated portrait and brief AI-written description; (D) hints: numbered AI suggestions with selectable buttons and ``Done''/``More Hints'' controls; (E) support tools: a table tracking story elements (``Who'', ``Where'', ``Objects'') under tabs for Help, Context, Keypoints; (F) story generator: an AI-written narrative snippet with two user-choice buttons to advance the plot and a button to use again the user input (A); (G) practice exercises: AI-generated prompts (``3 unusual pets...''; ``3 places to hide a treasure...'') with finish buttons and response fields.}
  \label{fig:teaser}
\end{teaserfigure}

%%
%% This command processes the author and affiliation and title
%% information and builds the first part of the formatted document.
\maketitle

\section{Introduction}
Improvisation is a vital skill in performance arts, requiring actors to think quickly, adapt to evolving contexts, and construct narratives in real-time. High-quality improvisation demands seamless integration of characters, plot points, and emotional responses, ensuring coherence and audience engagement. Achieving this balance involves significant cognitive effort and extensive practice. However, traditional improvisational theater (improv) training heavily relies on human interaction, limiting opportunities for individual practice outside structured group settings.

Advancements 
in artificial intelligence (AI), particularly large language models 
(LLMs) have introduced new avenues for supporting creative processes, including improv \cite{DBLP:journals/corr/abs-2405-07111, DBLP:conf/icccrea/MirowskiMBWVE20}. AI can generate story elements, suggest narratives, and simulate actor interactions. Current AI-based systems primarily focus on integrating AI as a supporting cast member or narrator in live performance \cite{DBLP:conf/aiide/MathewsonM18, DBLP:conf/aiide/MathewsonM17}. While these systems benefit from the random nature of LLMs, they often require \textit{immersion-breaking} human intervention% or control
, and struggle to keep \textit{narrative coherence} (i.e., consistent plot threads)  \cite{DBLP:conf/icccrea/BranchMM21, DBLP:conf/aiide/MathewsonM18}. 

\systemName shifts the focus from live performances to \textit{actor training} by leveraging AI to provide \textbf{automated, coherent, and creative support} for improvisation. Based on a formative study with 15 actors of different experience levels, we outline key challenges and needs, revealing ways AI can enhance improv practice. Our system uses \textit{multimodal input} -- actor's audiovisual performance -- and generates characters, objects, and scenarios, mimicking the \textit{unpredictability} of live performance, while reducing cognitive load by \textit{tracking narrative elements} throughout the improv session. Furthermore, \systemName incorporates\textit{ structured exercises} based on training techniques used by experienced actors, uses LLM randomness to support creativity \cite{DBLP:conf/hci/LiLS24, DBLP:journals/corr/abs-2410-03703}, and supports traditional improv practice. 
By requiring no additional equipment for speech and motion recording, the system %mimics live practice to align with actors' preference for traditional methods.
aims not to increase overhead for the actors and keeps the interaction seamless and intuitive.

In summary, our contributions include:
(1) A formative study with improv actors to gather insights on opportunities for AI-driven improvisation tools,
(2) \systemName, an AI-powered system for improv training with multimodal input, and 
(3) a pilot study with three experienced improv actors to evaluate \systemName.

\section{Related Work}

\subsection{AI in Improvised Theater}
% \cameraready{Traditional improv is a live, collaborative process where actors co-create narratives on stage, guided by audience suggestions.}
Traditional improvisation unfolds in real time, with actors responding instantly to each other's words and gestures. Performers often use verbal suggestions from the audience to shape pacing, direct the narrative, and introduce randomness.

Recent innovations have brought technology onto the improv stage. For instance, \textit{Improv Remix} uses mixed reality to let performers revisit recorded scenes and explore narrative possibilities \cite{DBLP:conf/ACMdis/FreemanB16}, and \textit{RIPT} which enlists an audience-controlled robot as a co-performer in live performences \cite{DBLP:conf/ACMdis/WunMTHO18}. While these works offer new modes of interacting with the performance itself, using AI can deliver a higher level of creative support for performers.

Early systems introduced AI to generate narrative elements and 
%spontaneous 
twists but required human guidance, lacking the automation needed for coherent storytelling \cite{DBLP:conf/aiide/MagerkoDD11, DBLP:conf/candc/MirowskiM19, DBLP:conf/aiide/MathewsonM18}. ~\citet{DBLP:conf/icccrea/BranchMM21} and \textit{Improbotics}~\cite{DBLP:conf/aiide/MathewsonM18} highlighted similar issues, where actors often compensated for AI inconsistencies, turning errors into comedic opportunities.
Advancements in LLMs have improved creative input, allowing AI to generate rich dialogue and interact more naturally \cite{DBLP:journals/corr/abs-2405-07111, DBLP:conf/fat/MirowskiLMM24, dogan_augmented_2024}. However, challenges persist, including biases in training data \cite{DBLP:conf/icccrea/MirowskiMBWVE20}, limited contextual awareness \cite{DBLP:conf/aiide/MathewsonM18}, and difficulties in maintaining narrative coherence \cite{DBLP:conf/icccrea/BranchMM21, DBLP:conf/aiide/MathewsonM17}. Cross-language improvisation has also been explored, as in ~\citet{DBLP:conf/icccrea/MirowskiMBWVE20}, revealing new possibilities but raising concerns about cultural and contextual relevance. 

Our system -- built on traditional improv techniques suggested by actors --  shifts AI's role from the stage to the backstage and focuses on supporting %improv
solo training rather than participating in live performances.

\subsection{AI-assisted Interactive Storytelling}
Interactive storytelling is increasingly achievable through modern technologies, enabling creators to build immersive, engaging narratives using tools like AR, IoT and more. 
Early systems -- like \textit{StoryMakAR}~\cite{DBLP:conf/chi/GlennICPR20} and \textit{Jigsaw}~\cite{DBLP:conf/chi/ZhangKCRTVM24} -- focused on interactive plots with these devices but struggled with accessibility due to sensory overload or technical challenges.
AI has further advanced storytelling by enabling dialogue-based narratives \cite{DBLP:journals/corr/abs-2011-04242, DBLP:conf/iui/ClarkRTJS18, DBLP:conf/iui/YuanCRI22, DBLP:conf/chi/ZhangXWYRWYWL22}. Tools like \textit{Crafting Narratives}~\cite{balasubramani2024crafting}, \textit{StoryDrawer}~\cite{DBLP:conf/chi/ZhangYWLLYY22}, and \textit{DrawTalking}~\cite{DBLP:conf/chi/RosenbergKWXP24} integrate diverse user inputs such as objects, drawings, or text to enhance creativity \cite{mystoryknight, DBLP:conf/acmidc/ZhaoCDFCLC24}. Collaborative storytelling, as seen in \textit{SAGA}~\cite{DBLP:conf/cscw/ShakeriND21, DBLP:conf/eacl/SwansonMPCD21}, positions AI as a co-creator. Despite these advancements, challenges like usability, scalability, and user adaptability remain \cite{DBLP:journals/expert/CavazzaCM02, DBLP:conf/aiide/ThueBSW07, yang2023storytelling, bowman2024immersive, DBLP:journals/tog/LiLHY22}.

Our system builds on these advances by positioning AI as a co-creator, focuses on narrative collaboration, and provides visual feedback via AI-generated images. Designed to be easy to use and flexible, it adapts to the needs of all users, with our AI that acts as an improv partner, using hallucinations to stimulate creativity.

\subsection{Performance-Based Motion Storytelling }

Motion-based storytelling systems use gestures or actions to direct narratives \cite{DBLP:conf/vr/SinghKHPGMR21, DBLP:conf/uist/WangQHHCR21, DBLP:conf/vr/ChenPS24}. Systems like \textit{Puppet Narrator}~\cite{DBLP:journals/jvca/LiangCDCTZ17, DBLP:conf/vsgames/LiangCKZJ15} and \textit{Ready...Action!}~\cite{DBLP:conf/ACMace/ChuQS14} engage users by capturing movements through markers or motion capture suites. While effective in enhancing interactivity, these systems are constrained by predefined gesture sets, reliance on costly equipment, and limited accessibility. For instance, \textit{Ready...Action!} allows children to act out scenes in real-time but requires external hardware and long setup times, reducing scalability and usability.

Our system takes advantage of recent advancements in visual understanding of LLMs to 
%allow for more accessible usage of real performance through motion understanding. 
analyze actors' performances using improv principles identified in our formative study.
This approach eliminates the need for specialized equipment or time-consuming setup, resulting in greater ease of use and accessibility.

\section{Formative Study}

In preparation for designing \systemName, we conducted a formative study with performers from an improv club. 
The first author conducted the analysis using a bottom-up approach \cite{braun2006UsingThematic}, with iterative refinements made during collaborative meetings with co-authors.
We collected 13 online survey responses and interviewed 2 participants, offering diverse perspectives on the integration of AI into improv. Participants varied in gender (F: 5, M: 10), age (18-25: 6, 25-35: 5, 35+: 4), and practice (< 1 yr: 3, 1-3 yrs: 10, 3-5 yrs: 1, > 5 yrs: 1).
% \paragraph{Role of AI in Improvisation}
% Participants expressed varied opinions on AI in improv training. 
A generational divide emerged: all younger participants (18-25) embraced AI, compared to 75\% (age 35+) - 80\% (25-35) of older ones. 
Many expressed AI's adaptive ability to provide \textit{immediate feedback} and \textit{personalize training}, while some -- based on experience with chatbots as \textit{ChatGPT} -- viewed AI as a substitute for human partners. Although AI as an on-stage actor was less favored, participants suggested that AI could simulate audience input by offering suggestions similar to reactions from a live crowd, despite concerns that over-reliance might undermine spontaneity.

\paragraph{Randomness and Narrative Coherence}
In many fields, AI hallucinations and randomness can be a problem, but not necessarily in the creative field. Many participants found that \textit{unexpected} stimuli (new characters, objects) can push them out of their comfort zone, facing unfamiliar situations.
Overwhelmed by these twists, participants highlighted the importance of structured \textit{key points} to guide the performance (settings, characters, plot elements). Even so, balancing narrative coherence and reactivity remains crucial: 46\% of participants preferred reactivity, 39\% were more neutral and 15\% favored narrative coherence. Although there is a slight overall preference for reactivity variability indicates that some actors favor spontaneity, while others prefer a more structured and rational approach.

\paragraph{Methods to Support Improv}
We discussed various approaches for supporting actors, including methods for tracking narrative elements, integrating images, and more.
When asked about tools for tracking these elements, 62\% were neutral and 13\% positive. In contrast, \textit{AI-generated images} for the story context received 70\% negative ratings due to ethical concerns, such as copyright issues for artists. Exercises aimed at \textit{training specific skills} were rated neutral (62\%) or positive (13\%), with many citing activities like \textit{finish stories suddenly} and \textit{quick-response challenges} to improve reactivity.
These findings, based on a small sample, highlight the need for broader research.

\subsection{Design Goals}
The study showed that actors are open to using AI as a training partner or audience-like support, where random hints can boost creativity. They also stressed the need for structured elements to track narrative details and exercises to train specific improv skills. 
Based on these results,
we propose four design goals (DG) for an AI-based improv support tool:

\textbf{DG1}:
Simulate an acting partner, facilitating traditional experience and allowing solo practice.

\textbf{DG2}:
Emulate audience interaction during improvisation, stimulating creativity through unexpected cues.

\textbf{DG3}: Help actors overcome difficulties in maintaining narrative coherence.

\textbf{DG4}: Provide realistic improv exercises, allowing actors to choose the skill to practice.

%-------------------------------------------------------------------------

\begin{figure*}[t]
    \centering
    \includegraphics[width=0.65\textwidth]{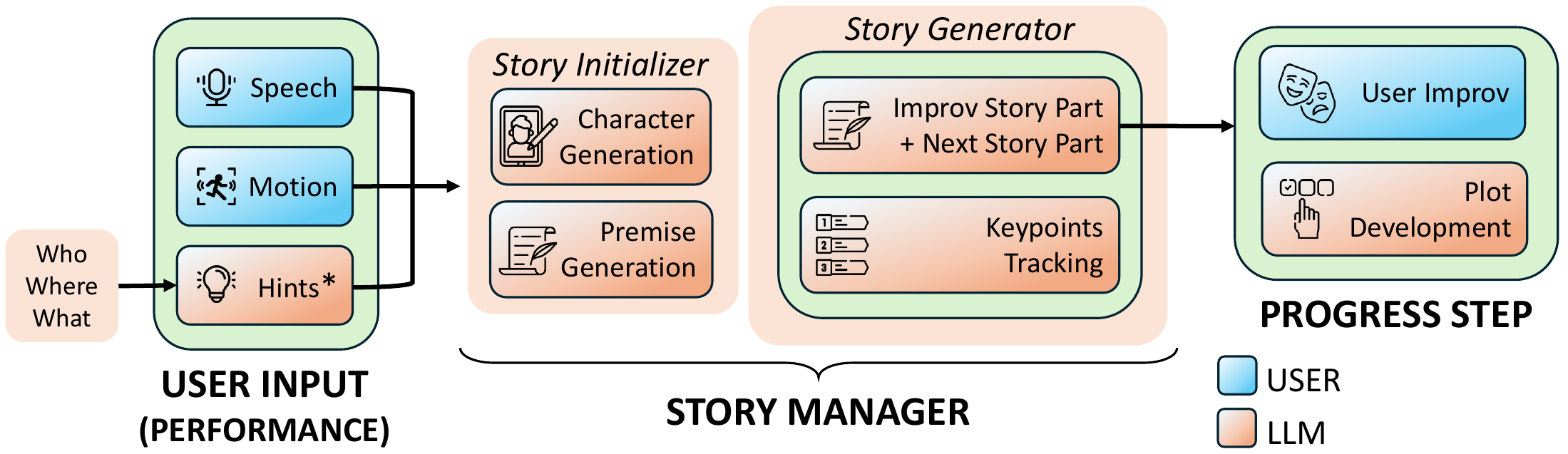}
    \caption{Single step of execution: the actor stages an improv performance (\textit{User Input}), using LLM-generated hints. The \textit{Story Initializer} creates the character and premise for the story. Next, the user's dialogue and movements are analyzed to continue the story through the \textit{Story Generator}. At this point, the user can choose whether to continue with another improv (return to \textit{User Input}) or continue with LLM-generated plot developments (return to \textit{Story Generator}). The possibility of concluding the story is also given.}
    \Description{Single execution step in the ImprovMate system. A left-to-right flow diagram in three stages: User Input (Performance), Story Manager and Progress Step. User Input contains ``Speech'', ``Motion'' and ``Hints'' feed into the system. ``Hints'' includes ``who'', ``where'', and ``what''. Story Manager is composed by Story Initializer and Story Generator. Story Initializer contains the character profile and story premise. Story Generator includes the current improv story segment, the next segment and key points tracking. Progress Step offers a choice between continuing with user-driven improv or invoking AI-driven plot development.}
    \label{fig:SingleStep}
\end{figure*}

\section{System Overview}
Based on these goals, we designed \systemName, a system that can support actors in their improv performances, recognizing dialogues and movements through LLM, and acting as a partner in the construction of the narrative.
Our goal is to enable practice without human partners, offering an exciting alternative to traditional improv while enhancing the experience with our AI-driven tools.

\subsection{System Implementation}
The system is implemented using \textit{React.js} for the frontend and \textit{Flask} for the backend.
\textit{OpenAI GPT-4o} \cite{DBLP:journals/corr/abs-2410-21276} is used to analyze user improvisations and generate coherent responses following improv principles; the prompts we used can be found in the Supplementary Material.
Insights from our formative study revealed three core actor needs: expressing themselves through motion and speech, co-creating with a scene partner, and receiving real-time support. These needs shaped our modular system design, resulting in three main components: \textbf{Performance Input}, \textbf{Improv Partner} and \textbf{Improv Support}, all integrated into the execution flow (\autoref{fig:SingleStep}).

\paragraph{Performance Input}
User performance is recorded via webcam and audio, and at each step the user may select AI-generated hints to incorporate into the scene.
After the user performs an improv scene, we analyze motion by sampling video frames at a rate of 1\textit{FPS}, sufficient for motion labeling using \textit{GPT-4o}~\cite{lin2023motionx, shotvl}, as determined in our interrogative study (see Supplementary Material for details).
The frames are sent to \textit{GPT-4o} for motion analysis. The audio is sent to \textit{OpenAI Whisper} for transcription. These together are used by the \textit{Story Manager} to advance the story. 

\paragraph{Improv Partner}
We defined the \textit{Story Manager}, an LLM agent divided into two components: the \textit{Story Initializer}, responsible for generating the premise, character, and beginning of the story, and the \textit{Story Generator}, which continues and evolves the story in line with the improvisational flow.
Initially, the Story Initializer is executed; then, for each step, the Story Generator uses the story so far to continue with the narrative. 
The latter is randomly influenced on the length of the text to generate and follows the principles of improvisation mentioned by formative study participants. It acts like a theater companion (\textbf{DG1}), introducing variables such as: new characters, new objects, changes of location, plot twists or time jumps.
After reading the generated story segment, the user continues either with a new improvised performance or by letting the AI further develop the narrative.

\paragraph{Improv Support}
Following the suggestions made by the formative study participants, we implemented several tools to support improv practice. \systemName tracks key points in the story (characters, places, and objects) to facilitate the maintenance of narrative coherence (\textbf{DG3}) and features optional audio narration that reads the story aloud. To simulate the audience's help and enhance creative thinking (\textbf{DG2}), the system proposes LLM-generated hints on %that the user can use. By exploiting the LLM's randomness, we generate suggestions on three key points that are fundamental for the performance: 
\textit{who}, \textit{where}, and \textit{what} they are doing. The user can regenerate suggestions or propose more scenarios for personalized improv practice.

To provide further support, \systemName provides two exercises 
%inspired by those mentioned by the actors, 
to refine the actor's skills (\textbf{DG4}).
In ``\textit{Three Things}'', the system proposes creative prompts (e.g., ``Three things you would take to the moon...'') which the user must complete with three responses, helping them train quick thinking and reactivity.
In ``\textit{Endings}'', the system first generates a story for the user to complete, then offers hints for different finale types (happy, sad, catastrophic, or absurd), challenging the actor to conclude the narrative in a single performance.

\begin{figure*}[t]
    \centering
    \includegraphics[width=1\linewidth]{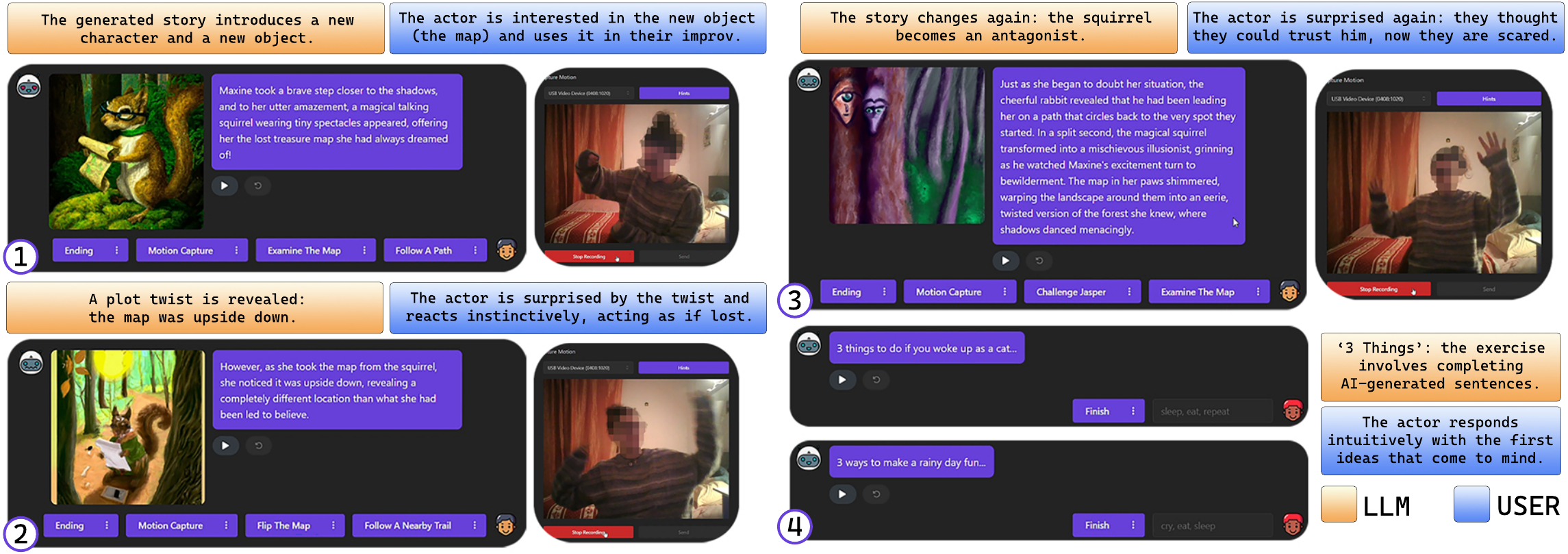}
    \caption{The experienced actor trying \systemName. Her reactions to what the story proposes are highlighted. (1-3) Actor using story mode. (4) Actor trying one of the exercises. Blue boxes refer to user input; orange boxes refer to LLM-generated content.}
    \Description{Experienced actor's responses across story and exercise modes. A four-part grid illustrating the actor's live webcam reactions (right) paired with AI-generated story parts (left), with orange callouts for large-language-model content and blue callouts for user actions: 1) the AI introduces a talking squirrel guide and a magical map; the actor examines the map on camera. 2) A plot twist reveals the map is upside down; the actor reacts with visible surprise. 3) The squirrel turns antagonistic in a twisted forest scene; the actor shows alarm. 4) In the ``Three Things'' exercise prompt, the actor rapidly supplies three ideas in the text field.}
    \label{fig:EvaluatedSystem}
\end{figure*}

\section{Pilot Study}
\label{evaluation}

To evaluate the usability and functionality of our tool, we conducted a pilot study with three improv experts (P1-P3) who participated in the formative study and were recruited from the aforementioned club. In this study, actors engaged with the main features of \systemName (\autoref{fig:EvaluatedSystem}).
Feedback was collected through observation, open-ended questions, and post-interaction interviews.
This study involved experienced improvisation actors participating in a low-risk, exploratory pilot of a training system. All participants were 18 or older, participated voluntarily, and were fully informed about the study's purpose and procedures. No identifiable or sensitive personal data were collected, and all interactions were recorded and analyzed in an anonymized form.

\paragraph{Story Generation}  
All participants were pleased that the AI was able to interpret their intentions, pauses and emotions, and generated a coherent narrative aligned with established character traits.
P1 appreciated that the AI added explanations to vague narrative elements she had introduced, adding depth to the story.
She described AI-generated improvisations as more ``discursive'' than traditional formats, providing richer context rather than focusing only on dialogues: \textit{``It's a different approach, but interesting [and] more reasoned''}. She explained that effective plot twists require complex story structures.
%and such tools can be useful for this purpose. 
Additionally, P3 appreciated the detailed descriptions that helped him imagine the scene.
The system's ability to resolve loose ends was seen as essential for high-quality performances (overcome difficulties, \textbf{DG3}).

\paragraph{Improv Support}
All participants found the AI-generated random hints and plot twists unexpected, interesting, and entertaining, similar to live suggestions from a director or audience (audience-like support, \textbf{DG2}). P1 quickly incorporated a newly introduced object into her improv, and was surprised by the LLM transforming a secondary character into a main one.
She appreciated being able to act ``freely'', choosing whether to react intuitively or thoughtfully, since a satisfactory outcome was guaranteed. 
P2 used the dynamic cues to enhance storytelling and suggested optional customization for the hints. P3 echoed these views and noted that the hints were instrumental in overcoming creative blocks.
The key point list, tracking narrative elements, reduced cognitive load by helping the participants focus on creativity rather than memory, recalling important details, as P1 said: \textit{``This was helpful, I always forget the names of characters introduced by others''}. P2 underlined the importance of the feature, emphasizing that constant visibility of it is crucial to avoid losing track of plot elements.

\paragraph{Exercises} 
The \textit{Endings} exercise helped participants practice narrative conclusions. P1 valued staging challenging scenarios (e.g., flash-forwards), while P3 appreciated refining endings through repetition.
\textit{Three Things} was described as fun and light, training rapid thinking with low cognitive load. P1 found it similar to live exercises, \textit{``like I do with my friends from improv.''}
P2 and P3 observed that structured exercises resemble live practice sessions, while free improv feels slightly different since AI replaces human interaction. 
The exercises were effective at training ``spontaneity and adaptability'' (P2), while requiring minimal effort, making them suitable for low-energy practice sessions (realistic exercises, \textbf{DG4}).

\paragraph{Further Benefits} 
The interface was described as ``intuitive'' (P2), with clear instructions and well-labeled AI-generated choices. 
P1 praised the audio narration, which helped stay focused on improvisation and P3 underlined the benefit for visually impaired actors.
P2 recognized the system's potential as both a creative companion and a teaching tool, while P3 noted its ability to build confidence in shy actors by letting them perform without audience pressure.
%thus fostering greater creativity.
All participants reported feeling entertained and creatively stimulated, as if \textit{``improvising with friends''} (replace traditional practice, \textbf{DG1}) (P1-P3).
\systemName accommodated various acting styles: P2 improvised with a much freer style, using different tones of voice and full character immersion, while P1 and P3 took longer to adapt to improvisation in front of a PC, possibly due to the lack of human companions and a theatrical atmosphere, as P2 noted.
Initially skeptical about AI in improv, \textit{``It is not easy to improvise in front of a PC [...] you could lose the atmosphere of the theater''}, P2 ultimately changed his mind, considering himself amused and satisfied by the experience.
A minor limitation was the occasional latency in generating AI responses, although the overall experience was enjoyable.

\paragraph{Comparison to Live Practice}
Participants found that \systemName efficiently replaces traditional theater training, it stimulates creativity through audience-style suggestions and new narrative directions, while enabling solo practice and realistic story creation. However, this approach lacks the real-time feedback that co-actors or teachers normally provide, limiting actors' ability to reflect and improve on their performance. Future work could introduce a guided-feedback feature to offer tailored critique and development paths; however, this will require deeper research into improvisational principles to ensure fidelity to the art form.

\section{Conclusion and Future Work}

We introduced \systemName, a system for a collaborative human-AI approach to performing theater, using dialogue and full-body motion as input to advance the narrative. It is designed specifically for improv actors. \systemName combines AI-driven storytelling with unassisted motion capture, enabling the creation of engaging performances without the need for additional external technology. Our pilot study has shown that actors are open and enthusiastic about integrating AI into their practice when the tools proposed are coherent with improv methods and resemble traditional approaches. %, appreciated by the participant.

In future studies, tracking creative progress over time could yield valuable insights into the benefits of our AI-based training system compared to traditional methods.  
Longitudinal studies can be conducted to understand the long-term impact on other aspects, such as cognitive abilities, creativity, or physical coordination.

%\section{ACKNOWLEDGEMENTS}

%%
%% The acknowledgments section is defined using the "acks" environment
%% (and NOT an unnumbered section). This ensures the proper
%% identification of the section in the article metadata, and the
%% consistent spelling of the heading.

\begin{acks}
This work was partially supported by the Japan Science and Technology Agency
(JST) as part of Adopting Sustainable Partnerships for Innovative
Research Ecosystem (ASPIRE), Grant Number JPMJAP2401.
\end{acks}

%%
%% The next two lines define the bibliography style to be used, and
%% the bibliography file.
\bibliographystyle{ACM-Reference-Format}
\bibliography{main}

%%
%% If your work has an appendix, this is the place to put it.
% \clearpage
\appendix

\section*{Supplementary Information}
We provide additional materials supporting the implementation and evaluation of our system. We include two sections:

\begin{itemize}
    \item \textbf{Motion Labeling Study}: we present a focused interrogative analysis of \textit{GPT-4o}'s ability to interpret and label motion-based input, which informs our understanding of its potential for multimodal improv interaction.
    \item \textbf{Prompt Design}: we detail the structured prompts used in the application to generate narrative elements and responses tailored to improvisational performance.    
\end{itemize}

These components offer insight into the technical decisions and empirical grounding behind our use of \textit{GPT-4o} in an improv training context.

\section{Preliminary Technical Evaluation of Motion Recognition from Video Frames}
\label{sec:appendixA}
By capturing user motion using a camera, we want to analyze the video to perform motion labeling. To achieve this, we performed an interrogative study on the motion labeling capability of \textit{GPT-4o}.
We run tests with different approaches using the \textit{Motion-X} \cite{lin2023motionx} dataset, which provides text and full-body motion annotations generated via a specialized pipeline that includes LLM. 

\subsection{Approach}

To perform an in-depth study, we want to analyze two different formats to represent motion in individual videos:
\begin{itemize}[leftmargin=0.5cm]
\item \textit{Video frames}: sampled frames inputted to the LLM for analysis.
\item \textit{MediaPipe data}: pose detection performed via MediaPipe, with results sent to the LLM.
\end{itemize}

Frames were sampled at varying rates to find a balance between detail and token usage. The LLM analyzed a subset of randomly sampled videos, comparing the generated descriptions to the ground truth labels of the dataset using the \textit{SentenceTransformer} model with BERT. Errors due to incorrect video recognition were excluded.

\subsection{Results}
The results show that the LLM accurately labeled motions even at low frame rates, e.g., 1 frame per second (fps), achieving comparable results at higher rates, e.g., 10 fps. The labels differed from the ground truth of the dataset, as the latter was generated with added context. Adding video titles as context improved similarity by about 10\%.
For each test, we computed the average similarity, the median, the standard deviation and the average tokens used for the three types returned as output by the request (completion, prompt, total), as shown in Tables 1 and 2.

For example, consider a video with a frame rate of 30 fps. With an \textit{fps\_skip\_ratio} of 0.1, frames are sampled every 3 frames (\textit{frames\_to\_skip = int(30 * 0.1) = 3}), resulting in more samples. On the other hand, with an \textit{fp\_skip\_ratio} of 1, frames are sampled every 30 frames (\textit{frames\_to\_skip = int(30 * 1) = 30}), resulting in fewer samples.
 Interestingly, the results demonstrate that fewer tokens (achieved by increasing \textit{fps\_skip\_ratio}, e.g., to 1) often lead to better similarity scores. This may be due to reduced redundancy in sampled frames, allowing the model to focus on essential motion patterns while avoiding noise and cognitive overload.

The tables report:
\begin{itemize}
    \item AVG (Average): The average similarity score between our labels and ground-truth.
    \item MED (Median): The median similarity score, showing the central trend in the data.
    \item STD (Standard Deviation): The variability of the similarity scores across all tested samples.
    \item AVG TKN CMP (Average Tokens - Completion): The average number of tokens used in the completion requests.
    \item AVG TKN PMT (Average Tokens - Prompt): The average number of tokens used in prompts.
    \item AVG TKN TOT (Average Tokens - Total): The total average tokens (combining completion and prompt tokens).
\end{itemize}

\begin{center}
    \begin{table*}[h]
        \centering
        \begin{tabular}{c|c|c|c|c|c|c}
             \toprule
             fps\_skip\_ratio & AVG & MED & STD & AVG TKN CMP & AVG TKN PMT & AVG TKN TOT \\
             \midrule
             0.1 & 0.581 & 0.593 & 0.165 & 88.010 & 10452.33 & 10540.344 \\
             0.25 & 0.584 & 0.605 & 0.155 & 88.358 & 4307.579 & 4395.937 \\
             0.5 & 0.603 & 0.599 & 0.147 & 87.896 & 2140.927 & 2228.823 \\
             0.75 & 0.598 & 0.586 & 0.154 & 86.621 & 1501.684 & 1588.305 \\
             1 & 0.602 & 0.624 & 0.160 & 85.958 & 1155.421 & 1241.379 \\
        \end{tabular}
        \caption{Results with video frames - different fps ratio. }
        \label{tab:my_label}
    \end{table*}
\end{center}

\paragraph{Table 1}
This analysis highlights the trade-off between sampling density (frames analyzed) and token efficiency, with higher skip ratios often leading to reduced token usage and close similarity scores.

\begin{center}
    \begin{table*}[h]
        \centering
        \begin{tabular}{c|c|c|c|c|c|c}
             \toprule
             fps\_skip\_ratio & AVG & MED & STD & AVG TKN CMP & AVG TKN PMT & AVG TKN TOT \\
             \midrule
             Context - 0.5 & 0.712 & 0.723 & 0.156 & 92.484 & 2147.979 & 2240.463 \\
             NoContext - 0.5 & 0.603 & 0.599 & 0.147 & 87.896 & 2140.927 & 2228.823 \\
             Context - 1 & 0.698 & 0.702 & 0.152 & 91.958 & 1173.698 & 1265.656 \\
             NoContext- 1 & 0.602 & 0.624 & 0.160 & 85.958 & 1155.421 & 1241.379 \\ 
        \end{tabular}
        \caption{Results comparison with or without context - fps\_skip\_ratio = 0.5, 1.}
        \label{tab:my_label}
    \end{table*}
\end{center}

\paragraph{Table 2}
The results show that adding video context consistently improves similarity scores across all settings, demonstrating its usefulness in enhancing motion recognition accuracy while not impacting token usage.

\begin{center}
    \begin{table*}[h]
        \centering
        \begin{tabular}{c|c|c|c|c|c|c}
             \toprule
             fps\_skip\_ratio & AVG & MED & STD & AVG TKN CMP & AVG TKN PMT & AVG TKN TOT \\
             \midrule
             Context - 0.5 & 0.712 & 0.723 & 0.156 & 92.484 & 2147.979 & 2240.463 \\
             NoContext - 0.5 & 0.603 & 0.599 & 0.147 & 87.896 & 2140.927 & 2228.823 \\
             MediaPipe - 0.5 & 0.441 & 0.417 & 0.164 & 125.763 & 11929.968 & 12055.731 \\
        \end{tabular}
        \caption{Results comparison video frames (with context and without) and MediaPipe - fps = 0.5.}
        \label{tab:my_label}
    \end{table*}
\end{center}

\paragraph{Table 3}
Although MediaPipe is a pose estimation tool, its use results in lower similarity scores, likely due to differences in motion representation or the overwhelming volume of data it generates, which also contributes to significantly higher token usage.

\section{Prompt Used}

Below, we provide some example prompts we provided to the LLM in the different parts of our system for replicability.

\subsection{Generate Story Improv}
This prompt is used to help the LLM continue the story based on user performance. It combines motion and audio with contextual information (character, setting, plot) to produce narrative text aligned with the improvisation.

% \begin{lstlisting}

\lstset{
  basicstyle=\small\ttfamily\setstretch{0.8}, % or try 0.85, 0.8
  breaklines=true
}

\begin{lstlisting}
You are a helpful assistant storyteller. Help me generate the next part of the story starting from the dialogue and motion performed by the actor.
1. The input object is a short improv performance that is to be used as a continuation for the narrative of the improvisation story.
2. Analyze the key movements in the video, focusing on how the performer's movements interact with their spoken words or sounds in the audio. 
3. Consider how these movements connect with the improvisational flow and transform or enhance the narrative in real-time.
4. As context for the improv performance, use the following hints to guide your analysis: {hints}. Be faithful to the context and use the characters, places or actions mentioned.
    - 'who': he character that the performer is impersonating, and will be the protagonist of the story.                        
    - 'where': Location where the story takes place.
    - 'what': Event used as the starting point of the story.
5. Consider the premise of the story, it consist main scenario or conflict of the story: {premise}.
6. Consider the narrative context that has been established so far: {story}.
7. Consider the key points that have been identified in the story: 
    - "who": The characters present in the story told so far.
    - "where": The location where the story takes place.
    - "objects": The objects present in the story told so far.
    - {keypoint}.
8. Generate the next story part based on the context of the story, considering description and emotion of the improv performance. The next story part should be: 
    - Not more than {length} sentences.
    - Take into account the "who", "where" and "objects" present in the story so far. Omit them only if they are not relevant to the new part.
    - Reflect any hints, decisions, or actions from the improv as part of the next story step.
    - Be true to the user's intentions.

Analyze the following improvisational performance with reference to the audio transcription: {transcription}

These are video frames in order: {video frames}
\end{lstlisting}
% }

\subsection{Generate Actions}
This prompt is used to provide users with actionable options that can meaningfully progress the story. Each action suggestion includes a concise title and a short description. These are used for selecting the user's next move, keeping the narrative interactive.

\begin{lstlisting}
You a great storyteller.
1. Understand the story so far.
2. Help me generate {number} unique actions the main character may perform.
3. Each action should advance the current story somehow.
4. Action is defined by:
    - Title, few words describing the action.
    - Description, very short paragraph with more details.
\end{lstlisting}

\subsection{Generate Story Part}
This prompt helps the LLM continue the story in a dynamic way, incorporating surprise elements such as plot twists, time skips, or absurd developments. It simulates the unpredictable nature of improv theater while respecting narrative consistency. The generation is guided by both the user's previous choices and random narrative settings.

% "Something absurdly good happens to the main character.",
% "Something absurdly bad happens to the main character.",
% "Introduce a new friendly character.",
% "Introduce a new relevant item.",
% "Advance the story in time (time skip).",
% "Move the story to a new location.",
% "Twist something already known.",
% "End with a cliffhanger.",

\begin{lstlisting}
You a great storyteller.
1. Understand the input object.
2. Understand the story so far.
3. Continue the story based on the main character performing the given action.
5. Generate a short visual description of a key moment in the new part:
    - Describe the environment.
    - Do not name the main character.
\end{lstlisting}

\subsection{Hints}
This prompt is used to generate creative hint combinations that serve as the initial inspiration for improv scenes. These story seeds provide context for the performer and the LLM, anchoring the narrative in a specific character, setting, and initial event.

\begin{lstlisting}
You are a helpful assistant. Help me generate some prompts to start an improvisation performance.
1. Generate {number} elements, each composed of 3 fields, the first answering the question 'Who?', the second 'Where?' and the third 'What happened?'
2. The answer to 'Who?' should be a character that can be used as a protagonist. (examples: a clown, a turtle, the Pope)
3. The answer to 'Where?' should be a location where the story takes place.
4. The answer to 'What happened?' should be a short event that can be used as the starting point of the story.
\end{lstlisting}

\subsection{Endings}
This prompt generates short, open-ended story beginnings. These openings are presented to the user as challenges to conclude in a single performance. 

\begin{lstlisting}
You a great storyteller.
1. Create an original story introduction which includes:
    - Character: Introduce a main character with a few unique traits.
    - Setting: Describe where the story takes place, incorporating vivid details.
    - Event: Describe an unusual or intriguing situation that the character encounters.
2. Develop the story so that it sets up a decision point or situation the character must respond to, without concluding the story.
3. Generate a visual description of a key moment in this part, capture the atmosphere and scene details.
\end{lstlisting}

\subsection{Three Things}
This prompt creates short, engaging questions that require users to quickly list three creative or humorous items or ideas.

\begin{lstlisting}
You are a creative assistant helping to design a fun, reactivity-based game.
1. Generate unique and engaging questions that prompt quick, imaginative responses. Each question should:
    - Be open-ended and encourage creativity.
    - Challenge players to think of "3 things" or similar sets, such as "3 items" or "3 ways."
    - Cover a mix of themes, including daily life, absurd scenarios, emotional situations, or unexpected events.
2. Questions should be short, clear, and easy to understand.
3. Ensure a variety of themes across the questions, such as:
    - Actions: "3 ways to climb up the stairs..."
    - Items: "3 things you would bring to the moon..."
    - Phrases: "3 things to say at a funeral..."
    - Emotions: "3 ways to show someone you care..."
\end{lstlisting}

\end{document}